# Bio-Soliton Model that predicts Non-Thermal Electromagnetic Radiation Frequency Bands, that either Stabilize or Destabilize Life Conditions

October 2016

J.H. Geesink and D.K.F. Meijer

Groningen, The Netherlands

**Abstract**

Solitons, as self-reinforcing solitary waves, interact with complex biological phenomena such as cellular self-organisation. Soliton models are able to describe a spectrum of electromagnetism modalities that can be applied to understand the physical principles of biological effects in living cells, as caused by electromagnetic radiation. A bio-soliton model is proposed, that enables to predict which eigen-frequencies of non-thermal electromagnetic waves are life-sustaining and which are, in contrast, detrimental for living cells. The particular effects are exerted by a range of electromagnetic wave frequencies of one-tenth of a Hertz till Peta Hertz, that show a pattern of twelve bands, if positioned on an acoustic frequency scale. The model was substantiated by a meta-analysis of 240 published papers of biological radiation experiments, in which a spectrum of non-thermal electromagnetic waves were exposed to living cells and intact organisms. These data support the concept of coherent quantized electromagnetic states in living organisms and the theories of Davydov, Fröhlich and Pang. A spin-off strategy from our study is discussed in order to design bio-compatibility promoting semi-conducting materials and to counteract potential detrimental effects due to specific types of electromagnetic radiation produced by man-made electromagnetic technologies.

**Introduction**

Extensive biophysical research has learned that typical discrete frequencies of electromagnetic waves can be favourable for living cells. An earlier analysis of 180 articles from 1950 to 2015, dealing with effects of electromagnetic waves on in vitro and in vivo life systems (Geesink and Meijer, 2016), has shown that discrete eigen-frequencies of electromagnetic waves are able to stabilize cells, whereas others cause a clear destabilization. In this paper we actualize these preliminary data by a further analysis of another 74 papers, with special reference to potential adverse effects of EM radiation, and detected a striking agreement with the earlier observed frequency pattern as depicted in Fig.2. On the basis of this research an obvious question arises: what can we learn from the existence of such (un)favourable EM (electromagnetic) radiation frequencies on living cells, realizing that the same frequency bands ranging from Hertz to sub-Terahertz are produced by man-made technology on a daily basis? Indeed, both living organisms and man-made technologies are able to generate electromagnetic pulses, that are transferred and processed at a non-thermal level. The main difference between both systems is that living organisms seem to be favourably affected by *coherent* patterns of electromagnetic waves, that may induce a "biological order", whereas modern man-made equipment not yet produces such selective coherent frequency bands. Our previous paper (Meijer and Geesink, 2016) was titled; "Phonon guided Biology", in the present publication we prefer the term Soliton as a further differentiation, since the latter term indicates a phonon/electron coupled quasiparticle formed through interaction with a lattice phonon cloud (solitons were also called polarons). Solitons are seen as electrically longitudinal vibrations that can travel along proteins, microtubules and DNA similar to semiconducting materials, inducing an endogenous electromagnetic field and interfere with local resonant oscillations by excitation of neighbouring molecules and macromolecules. Coherence is defined as the physical congruence of wave properties within wave packets and it is a known property of stationary waves (i.e. temporally and spatially constant) that enables a type of wave interference, known as *constructive*. Constructive wave interference leads to the generation of specific resonance patterns promoting coherent cellular domains and dynamic cell systems are partially operating via this principle. On this basis an important question arises: is it possible to increase the degree of coherency of man-made electromagnetic signals in order to externally support the health of living cells? Living cells have been demonstrated to be influenced by selective spectra of frequency bands in the UV, light, IR, THz and ELF range, as was consistently reported from many current bio-radiation research groups. Fröhlich proposed in 1968, that living cells, for constructive interference, seem to employ so called acoustic polarons, which can be described by Bose-Einstein-statistics. The particular wave-information that affects living cells at Terahertz frequencies, consists of oscillating charges in a thermal bath, in which a large numbers of quanta condense into a single state. The latter is known as a Bose-Einstein condensate, and induces a physical and non-thermal interaction between biomolecules. Bose-Einstein condensation represents a phenomenon wherein the bosons (particles/waves in quantum mechanics that follow Bose–Einstein statistics) produce a combination of waves that merges at the lowest cellular energy level into a shared quantum state. Such a collection of weak interacting waves, thereby constitute a discrete set of available energy states at a thermodynamic equilibrium. This results in a so called "macroscopic wave function", implying that all the particles/waves in the condensate take the form of a coherent wave modality. The latter can take wave dimensions that are many orders of magnitude larger than that of the microscopic objects such as atoms and molecules, due to the fact that the resulting overall wave field is highly correlated.

## 1. Coherence versus decoherence in life processes

Coherence or non-randomness of quantum resonances has also been discussed by Einstein and Infield (1961) for so-called "prequantum modes". It was Schrödinger who recognized that coherent interaction of waves is coupled to entanglement as 'the characteristic aspect of quantum mechanics' and suggested that "eigenstates"', also called "preferred states" are able to survive interaction with the environment. Coherent resonances can be present, for example, in electrons, photons and water molecules. The preferred locations receiving resonance transfer in the case of living cells, are the surrounding domains of ion water clathrates, nucleic acids and ion-protein complexes. These are present in and near neurons, ion channels, proteins and DNA. Water is known to be coherently nano-structured and coherent affecting bio-molecular processes, including protein stability, substrate binding to enzymes, as well as electron and proton transfer (Del Giudice, 2010, Chaplin, 2000, Johnson, 2009).

Quantum mechanics explains the interactions of wave/particles energy at the scale of atoms and subatomic particles. Quantum mechanics assumes that physical quantities such as energy or momentum are, under certain conditions, quantized and have only discrete values. Quantum coherence have been shown not only for micro states, but also for macro processes such as photosynthesis, magneto-reception in birds, the human sense of smell as well as photon effects in vision, all showing a non-trivial role for quantum mechanisms throughout biology (Rozzi, 2012, Lambert et al, 2013; Huelga and Plenio, 2013). Thus clear quantum coherence in living systems has been proposed (Swain, 2006). Apparently, nature makes use of wave information to induce and stabilise biological order. The stability and life times of these waves depend upon the extent of thermal decoupling of the stable state(s) of cells from the heat bath.

Yet, in order to maintain stability of bio-molecules in living systems, also *external* coherent information is required. Whether or not (externally applied) coherent electromagnetic radiation in the Terahertz region can effectively pump bio-molecules in and/or out of the meta-stable state(s) in relation to collective biochemical reactions, is a question of critical importance to the existence of associated bio-effects (Illinger, 1981). Interestingly, it is known that coherent wave information can be perturbed by de-coherent wave information and the reversed process can also occur in a kind of coherence-decoherence state cycling. Shor (1997) proposed a quantum error correction theorem for quantum calculations. If degrees of freedom in quantum computing are de-cohering due to loss of phase information from the computing system to its environment, the addition of coherent information to the system from the outside, turns the de-coherent state into a re-coherent one again. In relation to this, Kauffman examined data on photosynthetic systems (reviewed by Arndt, 2010,

Lloyd, 2012). In photosynthesis, photons are captured by the chlorophyl molecule that is held by an antenna protein, by which the chlorophyl molecule maintains quantum coherence for up to 750 femtoseconds. This is much longer than the classical prediction, and is viewed as responsible for the highly efficient energy transfer in photosynthesis. The particular antenna protein plays a role in preventing de-coherence, and in inducing re-coherence in de-cohering parts of the chlorophyll molecule. Kauffman proposed that this raises the possibility that domains of quantum coherence or partial coherence can also extend across neurons in the brain.

## 2. Coherent vibrations in the relation to Bose-Einstein condensates and suppression of anharmonicity

When in nature coherent states are induced to be organized and stabilized at the lowest possible energy level, then atoms, electrons, photons, bosons and magnons will potentially be ordered in the same way. The processes that mediate such events in biological systems are driven by fine tuned molecules, structured at the nano- and sub-nanoscale. At these small scales, the local dynamics are governed by the laws of quantum mechanics, as shown for photosynthesis in which excitons exhibit a wave like character. Hidalgo (2007) argued that the maximum payoff for a quantum system is provided by its minimum energy state. The individual system components will rearrange themselves to reach the best possible state for the whole system, being a sort of microscopic cooperation between quantum objects.

If energy is fed into these vibrational modes, then a stationary state will be reached in which the energy content of the electromagnetic modes is larger than in the thermal equilibrium. The excess of energy is supposed to be channelled into a "macroscopic wave function", like Bose Einstein condensates, provided the energy supply exceeds a critical value. Thus, under these circumstances, a given supply of energy is not completely thermalized but used in maintaining coherent electromagnetic states in living organisms. Fröhlich inferred that such orderings principles employ boson-like quasi-particles in biological processes, similar to condensed inorganic matter (semiconductors). He claimed in 1968 that oscillating charges in a thermal bath, in which a large numbers of quanta condense into a single state, form a condensed set of oscillators that can activate a vibrational mode of the lowest possible frequency at room temperature. Bose-Einstein condensation, in this manner, serves as a method for energy storage as well as for channelling energy to specific bioprocesses such as cell division and macromolecule synthesis. He considered the many boson system to be consisting of polar vibrations of biopolymers under excitation (longitudinal phonons due to metabolic energy pumping) that are embedded in a surrounding fluid. He derived a set of equations that describe the time evolution of collections of these vibrational modes in various model structures such as proteins or membranes. He also estimated, on the basis of theoretical arguments, that collective vibrational modes of metabolically active biological systems are in the frequency range of 0.1–10 THz (0.1–1.0 ps), displaying typical spectral features and that polarons play a crucial role in such processes (Fröhlich, 1968, 1969, 1988). *Polarons (also called Solitons, see later)* are quasiparticles, produced by coupling of electrons and phonons, producing collective excitations of surrounding atoms. The term polaron was earlier coined by Landau in 1933 to denote this type of quasiparticle, comprised of a charged particle coupled to a surrounding polarized lattice in the presences of photons. Next to *polarons*, *polaritons play a role, which* represent phonons coupled with excitons (electron hole pair that arise from excitation of electrons by photons, see Fig. 1). Fröhlich proposed a Hamiltanion model for polarons, through which their dynamics can be treated quantum mechanically as acoustic phonons located in a Bose–Einstein condensate (BEC).

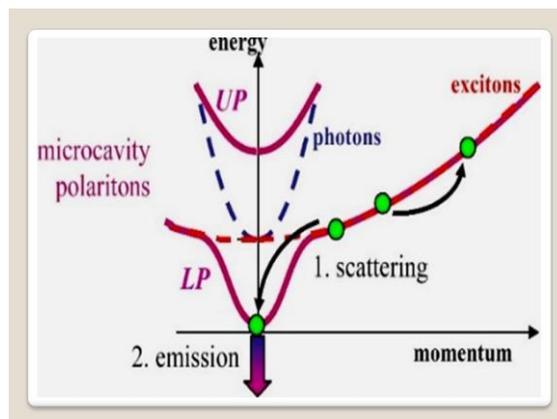

*Figure 1. Polariton dispersions (upper, UP, and lower, LP) are formed by a strong coupling between excitons (dashed red) and photons (dashed blue line). Polaritons (green circles) accumulate in the LP branch through stimulated pair scattering (arrows), before escaping radiative from the bottom of the LP branch (Högersthal and Baumberg, School of Physics University of Southampton)*

In the same time Davydov discovered the related principle of longitudinal wave forms called *solitons.* A soliton is defined as a self-reinforcing solitary wave that travels at constant speed without changing shape. He proposed that solitons play a role in the energy transfer and conformational states of biomolecules. The Davydov model, describes the interaction of the amide vibrations of peptide groups with the hydrogen bonds that stabilize the α-helix of proteins. The excitation and deformation processes balance each other and thereby form a soliton. His theory showed how a soliton could travel along the hydrogen bonded spines of the alfa-helix protein molecular chains (Davydov, 1973). He stated that solitons are able to suppress anharmonicity (the deviation of a system from being a harmonic oscillator) by the excitation of high quantum levels, a process that facilitates the crossing of potential barriers and the transfer of a molecule to a new conformational state. His concept of the excitations of atoms and solitons, as the quanta of collective vibrational motions of atoms and molecules in living cells, is also known from solid-state theory. The resulting excitation (excitons) moves through the protein uninhibited, much the way electrons move in a superconducting state. The excitons are bound states of electrons and electron holes present in semiconductors and liquids, which are attracted to each other by electrostatic Coulomb forces. Excitons interact with the lattice vibrations, under the action of electromagnetic fields and show a bound state. A bound state in quantum physics describes a system where a particle is subject to a potential such that the particle has a tendency to remain localised in one or more space regions. The energy spectrum of the set of bound states is discrete, unlike the continuous spectrum of free particles and shows typical *eigen-frequencies,* at which a system tends to oscillate on its own, in the absence of any driving or damping force. Davydov introduced a mathematical model to show how solitons could travel along the three spines of hydrogen-bonded chains of proteins. Davydov's Hamiltonian is formally similar to the Fröhlich-Holstein Hamiltonian for the interaction of electrons with a polarizable lattice (Davydov, 1977).

Thirty years later, solitons are a widely observed physical phenomenon that behave like waves but possess many features of particles (Lakshmanan; 2011). In biology, soliton theory has been applied to explain signal and energy propagation in bio-membranes as occurs for example in the nervous system, and to low frequency collective motions in proteins and DNA. Solitons do not obey the superposition principle, which makes the wave structure robust in collisions with other wave structures (Kuwayama and Ishida; 2013). Pang recently improved Davydov's model by incorporating a change in both the particular Hamiltonian and the intrinsic wave function of the system, proposing a quasi-coherent two-quantum state (Pang, 2016). It remains to be established whether solitons are also involved in higher-level biological phenomena.

## 3. A quantized acoustic scale able to predict frequencies that stabilize or destabilize living cells

According to Shapiro (1961) the concept of coherence in the field of acoustics only means something on an intuitive level, yet a formal definition has not been generally agreed upon. He proposed that when the particular processes can be called highly coherent if the variability of the phase differences between the signals is small, whereas if the processes are defined as incoherent if the phase difference has a high degree of variability. Interestingly we found series of coherent quantized frequencies located in 12 bands, which could be depicted on an acoustic reference scale: 256.0, 269.8, 288.0, 303.1, 324.0, 341.2, 364.7, 384.0, 404.5, 432.0, 455.1, 486.0 Hz, as previously published by Meijer and Geesink (2016). This not only agreed with the acoustic scale calculated by Ritz of eigen-frequencies of thin vibrating square membranes, but was also inferred from the basic frequency G-tone, positioned at 96 Hertz as measured for thin vibrating membranes by Chladni, both showing stable and coherent geometry patterns at discrete frequencies of this acoustic scale, which give rise to self-organisation. Additional frequency scales for electromagnetic wave values could be derived from this reference scale by multiples of $2^n$ (n is an integer); see for further details of the construction of the acoustic scale the appendix in Meijer and Geesink (2016).

The mathematician Ritz (1909) found the same intervals by computing eigenstates and eigen-frequencies for sound frequencies of vibrating thin square plates, such as discovered in the membrane studies of Chladni, 1787. Instead of trying to solve the partial differential eigenvalues directly, Ritz used the principle of energy minimization, from which the essential equations could subsequently be derived. He succeeded to calculate the different coherent geometric Chladni patterns and the first 35 overtones of these patterns, which could be duplicated in subsequent studies by others from 1970-2013. In this way a quantized acoustic frequency scale has been calculated based upon 12 scalars with a qualitatively small variability of phase differences and describing coherent geometric patterns (see for further details Geesink and Meijer, 2016, Meijer and Geesink, 2016).

We subsequently questioned whether biological systems are also sensitive to selective coherent electromagnetic wave frequencies, in which the variability of phase difference between the signals is small. Therefore, we performed a meta-analysis to validate published frequency data of biological studies in which cells and living organisms have been exposed to non-thermal electromagnetic waves. In total 114 different frequencies employed in 145 independent published studies covering a broad area of health effects were identified able to stabilize living cells in vivo and in vitro (called beneficial effects of EM radiation, see Appendix 3).

All measured beneficial biological data of the applied frequencies in the Hz, kHz, MHz, GHz and THz range could be accommodated by the single reference acoustic scale, ranging from 256 till 486 Hz. All the values of the "beneficial frequencies" are indeed located in 12 distinct quantized frequency bands. The different measured beneficial biological data fit precisely with the calculated stabilizing frequencies (see the green circles and the green triangles in Figure 2) and are positioned again and again at a typical stabilizing frequency, each within a mean bandwidth of 0.74% (see the green ellipses and for the deviations of the individual values from the theoretical values Appendix 1).

Also 60 different measured frequency-data were identified in 74 independent published biological studies, that rather seem to destabilize living cells (called detrimental effects of EM radiation, see Appendix 3). The different measured detrimental biological data fit with the calculated destabilizing frequencies "in between the beneficial frequency bands" (see the red squares and the red triangles in Figure 2) and are positioned again and again at a typical destabilizing frequency, each within a mean bandwidth of 0.88% (see the red ellipses and for the deviations of the individual values from the theoretical values Appendix 2).

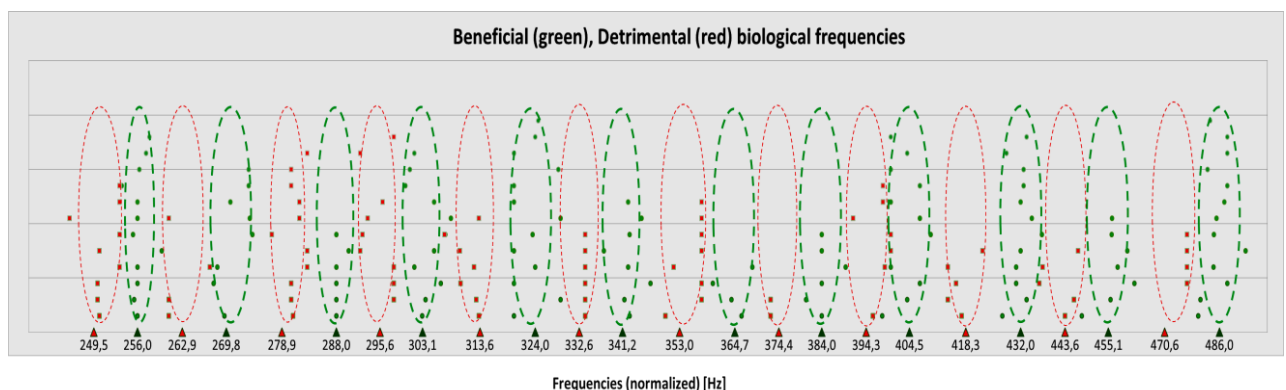

**Figure 2. Calculated normalized EM frequencies that were experimentally applied to life systems are found to be patterned in 12 apparent bands of life-sustaining frequencies (green points) and detrimental frequencies (in red) are positioned in between the beneficial frequency bands**

*(148 beneficial EM frequencies, positioned in 12 bands (green points) and 77 detrimental EM frequencies (red squares) in between the life-sustaining bands. Effects were measured following exposure of living cells in vitro and in vivo to the indicated (non-thermal) EM radiation conditions, as derived from 219 separate biological studies. The reported life sustaining frequencies (in Hz, kHz, Mhz, Ghz, Thz, Phz) appeared to be positioned in 12 discrete bands, if normalized by multiplying or dividing by $2^n$ (n is an integer) and plotted on a logarithmic acoustic scale, whereas the adverse frequencies (depicted in red) are shown to lay in between the beneficial frequency bands. Green points plotted on x-axis represent life-sustaining frequencies; Points in red represent the life-destabilizing frequencies. Each point indicated in the graph is taken from earlier published data and indicates a typical frequency for (a) biological experiment(s). For clarity, points are evenly distributed along the Y-axis)*

The zones, which are located in between the designated regions of stabilisation and destabilisation are probably transformation zones of geometric standing wave like patterns as shown by Chladni for acoustic patterns in membrane vibration experiments (1787), which have been replicated by independent researchers in studies from 1970-2013. The bandwidth of transformation regions of patterns (the mutual distance of the ellipses) is estimated at about 0.50% of the applied algorithmic frequency.

We propose to consider the identified 12 basic eigen-frequencies of the acoustic wave-function, as defined by the inferred frequency scale, for an *extended bio-soliton model*. This might be a complementation of the earlier models that were initially derived by Davydov and Fröhlich for a supposed "macroscopic wave function". This proposal is further supported by the calculated frequency data of Ritz (1909), the measurements of Chladni (1787). In such an approach, all of the registered beneficial frequencies for living cells can be integrated in an extended "macroscopic wave function" that includes a broad coherent frequency range of 0.1 Hertz – PHz (see the appendix 1 and 2).

A quantized acoustic scale might have been found to predict frequencies that stabilize or destabilize living cells and the experimentally applied frequencies turned out to be very close to theoretically calculated frequencies, which are positioned at typical algorithmic frequencies within a bandwidth.

*Conclusion: These observations provide clues for the existence of a specific pattern of electromagnetic radiation that potentially may affect the viability of life systems and may be involved in the functional structuring and self-organisation of bio-molecules within cells through organizing them at the lowest possible energy level.*

## 4. Physical models about biological influences on cells by electromagnetic waves

Research about electromagnetic pulses on living cells has been systematically undertaken the past eighty years. About 25.000 biological/physical reports are available, of which a part is dealing with non-thermal biological effects on cells. Influences of electromagnetic waves causing thermal effects on biological systems are relatively well understood, and more knowledge about non-thermal effects of electromagnetic waves have become available. Four basic physical models have been developed the past fifty years describing physical principles of non-ionizing radiation on biological effects of cells: 1) ion cyclotron resonance models, 2) resonant recognition model, 3) radical concentration models, and 4) models about the stability of coherent waves.

1) The models of Blackman (1985) and of Liboff (1985) describe the role of ion cyclotron resonances (ICR's) that involve a combined action of an Extreme Low Frequency (ELF) electromagnetic field and the static geomagnetic field on the resonance of typical ions. The model of Lednev (1991, 1993) proposes the ion parametric resonance hypothesis, which predicts that when a frequency of a combined dc-ac magnetic field equals the cyclotron frequency of an ion, for example calcium, the affinity of the calcium for calcium-binding proteins such as calmodulin will be affected. It is also considered that ions bonded to proteins ($Ca^{2+}$, $K^+$, and/or $Mg^{2+}$) behave as isotropic coupled oscillators. There is not yet consensus about the replication of the biological effects at the same typical frequency of combined static and time-varying fields under resonance conditions predicted by these models (Blanchard, 1994; Hendee, 1996; Zadhin, 2005; Halgamuge, 2009).

2) Cosic introduced the concept of dynamic electromagnetic field interactions and that molecules recognize their particular targets and vice versa by the principle of electromagnetic resonance: the Resonant Recognition Model (RRM). All molecules have their own spectrum of vibrational frequencies and DNA itself can function as an aggregate of EM antennae that could discern, differentiate, and transform EM energies to perturbations in protein sequences (Cosic, 1994). Periodicities within the distribution of energies of delocalised electrons along a protein molecule are crucial to the protein's biological function, i.e. interaction with its target. The RRM makes use of the equation of the Electron Ion Interaction Potential (EIIP), which is coupled to a theoretical modulated weak potential experienced by electrons in the vicinity of ions and the cloud of surrounding electrons. The RRM calculates indirectly the spectral characteristics of proteins at frequencies of infrared, visible light and ultraviolet (Cosic, 1997, 2007, 2015, 2016; Veljkovic, 1972; 1985; Pirogova, 2001; Dotta, 2013).

3) The models of Barnes and Greenebaum (2014), and Buchachenko (2016) picture that radical concentrations of biomolecules can be influenced by combinations of steady and alternating magnetic fields that modify the population distribution of the nuclear and electronic spin states at a relatively high magnetic field strength. It is biologically known that changes in free radical concentrations have the potential to lead to biological significant changes.

4) The polaron model of Fröhlich (1968) and the soliton model of Davydov (1973) are models that describe effects of coherent states of waves for inanimate as well as animate systems and interactions with electromagnetic waves, electrons, phonons, polarons, polaritons and magnons with regard to wave stability. Polarons are similar to solitons, and we will further use the name soliton. After all John Scott Russell in 1834 was one of the first to observe a soliton: a solitary wave in the Union Canal in Scotland, which maintained its shape while it propagates at a constant velocity, see Fig. 3.

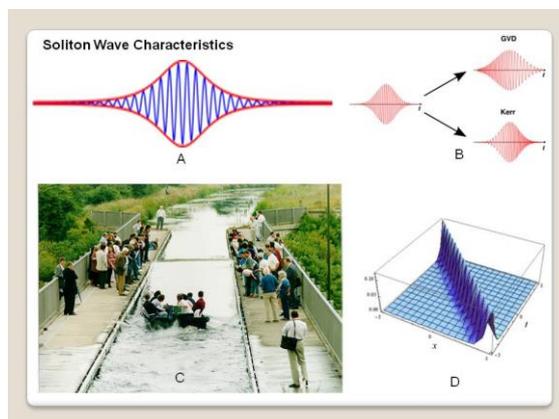

*Figure 3. A: Longitudinal soliton wave. B: Linear and non-linear soliton wave propagation C: Union Canal in Scotland as a spot of the discovery of the solitary wave. D: Optical soliton wave representation*

# 5. Potential support for a bio-soliton model

Many researchers have contributed to the soliton research and much progress has been made: Fermi, Pasta and Ulam (1955), Zabusky and Kruskal (1960-1970), Fröhlich (1968-1988), Davydov (1973-1991), Chou (1976-1994), Luzzi (1982-2012), Pokorny (1982-2015), Pang (1990-2016), Heimburg (2005-2016), Chin (2010-2013), Lundholm (2015), among others. The flow of soliton-like energy in a one dimensional lattice consisting of equal masses connected by nonlinear springs has been calculated (Fermi, Pasta, Ulam (1955). Porter showed later that the energy initially put into a longwave length mode of a system does not thermalize but rather exhibit energy sharing among the few lowest modes and long-time near recurrences of the initial state and would not transfer energy into higher harmonic modes (Porter, 2009). Zabusky and Kruskalz (1965) concluded that the equation of Korteweg and de Vries (1895) admits analytic solutions representing what they called solitons: propagating pulses or solitary waves that maintain their shape and can pass through one another, see Fig. 4.

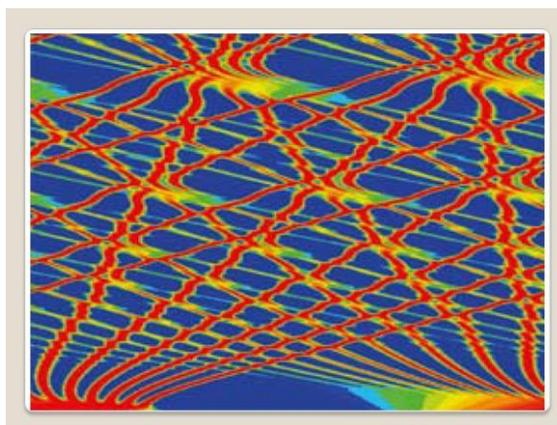

*Figure 4. A system like the one Fermi, Pasta and Ulam modelled, gives rise to solitons, which propagate in either direction, exchange positions and eventually return the system to states that resembles its initial configuration. The motion of the solitons can be seen here by following the lines of colours, which denote displacements (From Porter, 2009 and image from Zabusky, Sun and Peng 2006)*

Their analyses showed that the coherence of solitons can be attributed to a combination of nonlinear and dispersive effects. Chou found indications for a similar soliton principle in proteins, in which the conformational protein adaptation is influenced by low frequency phonons acting as an "information system" (Chou, 1977). More detailed experimental evidence for the existence of collective excitations of low-frequency phonons came available for proteins (Chou, 1985, 1994; Xie et al., 2002) and for polynucleotides (DNA and RNA) through the observation of low-frequency oscillations modes in the Raman and far-infrared spectra of proteins (Painter et al., 1981, 1982; Fischer et al., 2002).

Heimburg and Jackson (2005) argued that electromechanical solitons, with properties similar to those of an action potential, can travel along the nerve axons. The compressibility's of biomembranes seem to be nonlinear functions of temperature and pressure in the vicinity of the melting transition, and show the possibility of soliton propagation.

Sinkala (2006) calculated the soliton-mechanism of electron transport in alfa-helix sections of proteins, using classical Hamiltonian analysis. He confirmed that folding and conformation changes of proteins are mediated by interaction with solitons which propagate along the molecular chain. In fact, many biological processes in any living organism are seen to be associated with conformational changes, as a result of space propagation of energy and electrons along protein molecules. One example is the 0.42 eV energy, released under hydrolysis of ATP molecule, as studied by Davydov (1973) and Pang (2001). A hypothesis is that this energy is transferred along the alfa-helical protein molecules, while the oscillation energy of the C = O moieties of the peptide groups (amide-I vibration) is at 0.21 eV or 1665 cm$^{-1}$, which is in resonance with the 0.42 eV of the ATP process. Interestingly, the energy released under hydrolysis of ATP molecule, the oscillation energy of the C = O of the peptide groups and the different bending modes of interfacial water molecules fit precisely with the calculated scalar frequencies of the acoustic wave-function, respectively: 0.415 eV, 0.2073 eV and 1660-1693 cm$^{-1}$.

Srobar (2015) and Pokorny (1982) proposed a wave equation for a Fröhlich system for cellular physiology, which describes the coherence between individual oscillations, using a number of energy quanta concentrated in one vibrational mode above the thermal equilibrium level and suggested an ensemble of interactions characterized with three coupled oscillators, of which the oscillator eigen-frequencies fall into the MHz and the THz frequency domains.

Pang et al (2016) showed that the distributions of the quantum vibrational energy levels of the protein molecular chain are crucial and used the Davydov theory in relation to the nonlinear Schrödinger equation to describe the resonant behaviour of proteins (Pang, 2001). A system of particles or waves can be described in terms of an "energy functional" or Hamiltonian, which represents the energy of a proposed configuration of particles or waves. In this system certain preferred configurations are more likely than others, which is approached by an eigen-analysis. The Hamiltonians of the lattice vibration and the soliton wave function of Davydov have been refined by Pang. The equations thereby express the features of collective excitation of phonons and excitons caused by the nonlinear interaction. The interaction domain contains two vibrational quanta (excitons), while the energy spectra of proteins at infrared frequencies have been measured. The Davydov–Pang model is theoretically plausible and more appropriate to the alpha-helical proteins containing amide-I, and show lifetimes of solitons at 300 K (Su, 2011). Pang also confirmed that the soliton mechanism of energy transport in life systems can be disturbed by man-made electromagnetic fields. Using both equations he succeeded to show that external man-made electric magnetic fields are able to depress the binding energy of the soliton, decrease its amplitude, and change its wave form. The calculated suppression of the solitons could be experimentally validated by measuring the infrared spectra of absorption of collagens activated by external electromagnetic waves. The stability of the DNA structure was shown to be dependent on the continuous input of phonon guided oscillations at which a cloud of entangled electrons surrounding the nucleotides acts as a harmonic oscillator at room temperature.

Also pigment-protein complexes generate non-equilibrium vibrational processes that lead to sustenance of electronic coherence, at physiological temperatures (Chin, 2013). The measured coherence times of several hundreds of femtoseconds are long enough for excitation energy transfer and excitonic coherence to coexist. Based upon this knowledge a semi-classical model with discrete modes was proposed that protects oscillatory excitation energy transfer and electronic coherence against background decoherence.

Solitons are unstable in three dimensions and decay through snaking instability. Solitonic vortices were proposed to bridge the gap between solitons and vortices, and have elements of both. Like vortices, their phase swirls around the defect, but only in a small region of space. Outside this region, the phase becomes homogeneous, like the phase on either side of a soliton. The 3D-soliton can propagate with a constant velocity along a vortex core without any deformation (Chevy, 2014; Adhikari, 2015). Vortex solitons at the interface separating two different photonic lattices – square and hexagonal – have been demonstrated numerically. The conditions for the existence of discrete vortex states at such interfaces have been considered and a picture of different scenarios of the vortex solutions behaviour developed (Jović Savić, 2015).

Meijer and Geesink (2016) proposed that metal ions influence brain function and health like calcium, sodium, potassium, copper, zinc, iron, manganese, cobalt, and lanthanides. In fact, it has been estimated that half of all proteins in the body form complexes with metals. These ions are cofactors in a wide range of brain cell functions, including cellular respiration, cell signaling at synapses, antioxidant removal of toxic free radicals and oxygen delivery to brain cells. It is considered that ions bonded to proteins ($Ca^{2+}$, $K^+$, and/or $Mg^{2+}$) behave as isotropic coupled oscillators of which cyclotron resonances show typical discrete frequencies, which might be described by a soliton wave function. Calcium transport and protein/channel binding is largely affected by magnetic fields with extreme low frequency (ELF) signals. $Ca^{2+}$ takes a crucial position in the integration of the distributed information within the cell, and affects at least 10 different cellular processes that have been shown to correlate with modalities of conscious perception.

Lundholm (2015) found a direct experimental support for the Fröhlich condensation in the arrangement of proteins, by detecting Bose-Einstein condensate-like structures in biological matter at room temperature. The group used a combined terahertz measuring technique with a highly sensitive X-ray crystallographic method to visualize low frequency vibrational modes in the protein structure of lysozyme. Structural changes, associated with low-frequency collective vibrations, as induced in lysozyme protein crystals by irradiation with non-thermal 0.4 THz radiation, were detected. The vibrations were sustained for micro- to milli-seconds, which is 3–6 orders of magnitude longer than expected if the structural changes would be due to a redistribution of vibrations upon terahertz absorption according to Boltzmann's distribution. The influence of this non-thermal signal is able to changes locally the electron density in a long alfa-helix motif, which is consistent with an observed subtle longitudinal compression of the helix.

## 6. How can (de)stabilisation of cells take place at a non-thermal level?

Soliton models are able to describe cellular electromagnetism, but also show underlying physical principles of biological effects in cells caused by non-ionizing electromagnetic waves. Stabilisation of cell-states takes place when the "macroscopic wave function" applied in these models, is active. Discrete coherent Terahertz frequencies and much lower and higher frequencies (from some Hertz to PHz) are coherently coupled, and obey Bose Einstein condensation, which are able to stabilize living cells at and near a dynamic equilibrium around room temperature. The stabilisation of cell states will occur at typical discrete frequencies, described by the particular wave function, in which each type of cell or bio-molecule or part of the bio-molecule will have its own eigen-frequency. Solitons, confined in a harmonic potential, can actually reabsorb energy in periodic cycles (Parker, 2003). An impression of the band width of these particular stabilizing frequencies has been given by Geesink and Meijer (2016) and shown in Fig.2.

The knowledge initiated by Davydov and Fröhlich and the model of Pang showed that the soliton mechanism of energy transport in life systems can be disturbed by man-made electromagnetic fields (Pang, 2016). It is postulated that the decay of the "macroscopic wave function" can be initiated when the orderings principle of the boson-like quasi-particles is disturbed by a sufficient amount of de-coherent electromagnetic waves. Such de-coherent waves exhibit frequencies, which do not fit in the range of eigen-frequencies of electromagnetic waves, and are called dark solitons, see Fig. 6. Dark solitons have already been observed in optical systems, and in harmonically confined atomic Bose–Einstein condensates. If such a dark soliton propagates in an inhomogeneous background, it dissipates coherent energy. Our EM frequency data indicate that destabilizing frequencies (dark solitons) are located "in between the beneficial frequency bands" (see Fig.2), which are positioned in between some Hertz and Terahertz.

Cellular plasma water is generally supposed to act as a transfer medium for external electromagnetic waves to biomolecules. The cellular plasma exhibits a highly arranged 3-D geometric structure as a liquid crystal, that exhibits surface interactions with macromolecular structures. The absorption spectrum between 1 THz and 10 THz of solvated biomolecules is sensitive to changes in fast fluctuations of the water network. There is a long range influence on the hydration bond dynamics of the water around binding sites of proteins, and water is shown to assist molecular recognition processes (Xu and Havenith, 2015). "Biological water" supports itself by coherent dipolar excitations and terahertz/femtosecond infrared interactions and these dynamics extends well beyond the first solvation shell of water molecules. The reorientation of water molecules around ions and interaction with solvated ions slows down during external THz-waves, shifting the absorption peak to lower frequencies and thereby reducing the absorption of radiation at THz frequencies (depolarization) (Tielrooij and Bakker, 2009, 2010). The coherence of liquid water is affected by applied external electromagnetic waves much weaker than those allowed according to the kT threshold (Del Giudice, 2010) and the reorientation of water molecules might take place at energy levels of less than a hundred $mW/cm^2$, which is at a non-thermal energy level.

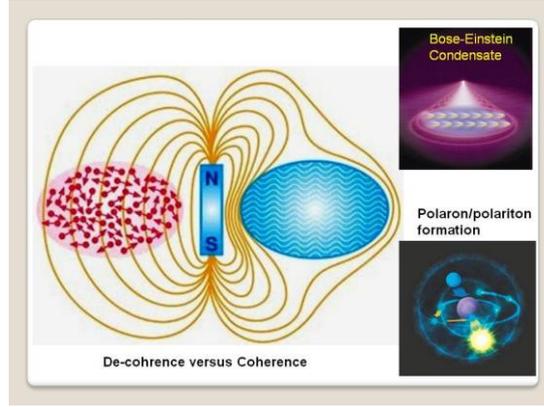

*Figure 5. Coherence Versus Decoherence, and Bose-Einstein Condensation*

Summarized: living cells make use of coherent frequency signalling, similar to Bose Einstein condensates, in order to stay stable. If incoming man-made electromagnetic signals exhibit a lack of coherency, than these signals potentially will decrease the coherency of (quantum) wave domains of living cells. A possible way to deal with such a problem is to lower the energy of the external man made waves and to add more coherent frequencies next to potential detrimental frequencies and by increasing the coherency of the electromagnetic signals through the use of appropriate semiconductor-technologies. Coherent Terahertz waves, obeying to the acoustic wave function, could be produced by appropriate semiconducting materials inserted in electromagnetic man-made devices, while making use of the so called Terahertz gap, enabling the combination of optical and electronic coherent information.

## 7. How to further constitute an integral bio-soliton mathematical model

The Fröhlich/Davydov concept has been elaborated and further improved by Pang taking into account that solitons can be largely stabilized, and their life-time increased due to mutual interaction of the particles with lattice vibrations. Consequently the total state of the system has been expressed in three different Hamiltonians. Due to this extensions, the solitons obtain life-times that are more compatible with the ruling biological conditions. This was expressed in Hamiltonians on quasi-coherent two-quantum state wave function (Pang, 2016). According to the mechanism of bio-energy transport Davydov established the theory of bio-energy transport in protein molecules, in which he gave the Hamiltonian of the protein molecules:

*Equation 1a:*

$$H_D = \sum_n [\varepsilon_0 B_n^+ B_n - J(B_n^+ B_n + B_n B_n^+)] + \sum_n [\frac{P_n^2}{2M} + \frac{1}{2}w(u_n - u_{n-1})^2]$$
$$+ \sum_n [\chi_1(u_{n+1} - u_{n-1}) B_n^+ B_n] = H_{ex} + H_{ph} + H_{int}$$

B is the creation (annihilation) operator for an Amide I quantum (exciton) in the site n, un is the displacement operator of amino acid residue at site n, Pn is its conjugate momentum, M is the mass of an amino acid molecule, w is the elastic constant of the protein molecular chain, N is a nonlinear coupling parameter and represents the size of the exciton-phonon interaction in this process, J is the dipole-dipole interaction energy between neighbouring amino acid molecules, m is the average distance between the neighbouring amino acid molecules. The wave function of the system proposed by Davydov has the form of:

*Equation 1b:*

$$D_2(t) = |\varphi_D\rangle|\beta(t)\rangle = \sum_n \varphi_n(t) B_n^+ \exp\{-\frac{i}{\hbar}\sum_n [\beta_n(t) P_n - \pi_n(t) u_n]\}|0\rangle$$

Pang has added a new coupling interaction of the excitons with the displacement of amino acid molecules into the Hamiltonian and replaced further the Davydov's wave function of the one-quantum (exciton) excited state by a quasi-coherent two-quantum state. The representations in equations (1a) and (1b) for the single-channel protein molecules are replaced by:

***Equation 2a:***

$$H = H_{ex} + H_{ph} + H_{int} = \sum_n [\varepsilon_0 B_n^+ B_n - J(B_n^+ B_n + B_n B_n^+)] + \sum_n [\frac{p_n^2}{2M} + \frac{1}{2}w(u_n - u_{n-1})^2] +$$
$$\sum_n [\chi_1(u_{n+1} - u_{n-1})B_n^+ B_n + \chi_2(u_{n+1} - u_n)(B_{n+1}^+ B_n + B_n^+ B_{n+1})]$$

***Equation 2b:***

$$|\Phi(t)\rangle = |\alpha(t)\rangle|\beta(t)\rangle = \frac{1}{\lambda}[1 + \sum_n \alpha_n(t)B_n^+ + \frac{1}{2!}(\sum_n \alpha_n(t)B_b^2)^2]|0\rangle_{ex} \times$$
$$\exp\{-\frac{i}{\hbar}\sum_n [\beta_n(t)P_n - \pi_n u_n|0\rangle_{ph}$$

Ritz (1909) had shown that an analytical calculation of acoustic eigen-frequencies of a thin square vibrating membrane is possible (Geesink and Meijer, 2016). Part of the particular calculation of Ritz is presented below (see Gander and Wanner, 2010, 2012):

Calculating ...
To evaluate $J(w_s)$, we thus have to evaluate

$$\int_{-1}^{1}\int_{-1}^{1}\left(\frac{\partial^2 w_s}{\partial x^2}\right)^2 = \int_{-1}^{1}\int_{-1}^{1}\left(\frac{\partial^2 \sum_{m,n} A_{mn} u_m(x) u_n(y)}{\partial x^2}\right)^2 dxdy$$
$$= \sum_{m,n}\sum_{p,q} A_{mn} A_{pq} \underbrace{\int_{-1}^{1}\int_{-1}^{1} \frac{\partial^2 u_m(x)}{\partial x^2} u_n(y) \frac{\partial^2 u_p(x)}{\partial x^2} u_q(y) dxdy}_{c^1_{mnpq}:=}.$$

Now $c^1_{mnpq}$ can be computed, since $u_n$ is known! Similarly

$$\int_{-1}^{1}\int_{-1}^{1}\left(\frac{\partial^2 w_s}{\partial y^2}\right)^2 dxdy = \sum_{m,n}\sum_{p,q} A_{mn} A_{pq} c^2_{mnpq}$$
$$\int_{-1}^{1}\int_{-1}^{1} 2\mu \frac{\partial^2 w_s}{\partial x^2}\frac{\partial^2 w_s}{\partial y^2} dxdy = \sum_{m,n}\sum_{p,q} A_{mn} A_{pq} c^3_{mnpq}$$
$$\int_{-1}^{1}\int_{-1}^{1}(1-\mu)\left(\frac{\partial^2 w_s}{\partial x \partial y}\right)^2 dxdy = \sum_{m,n}\sum_{p,q} A_{mn} A_{pq} c^4_{mnpq}$$
$$\int_{-1}^{1}\int_{-1}^{1} w_s^2 dxdy = \sum_{m,n} A_{mn}^2 \quad \text{(orthogonality!)}$$

We found that the calculated eigen-frequencies by Ritz are fully compatible with the collected beneficial frequency data for living cells that we inferred from the earlier mentioned published biological studies (Meijer and Geesink, 2016). In this paper we therefore would like to propose a combination of the knowledge of Pang (2016), and the frequency scale of Ritz (1909), which could be validated by many biological data from 1966 till now (Fig.2).

***Summarizing***

The bio-soliton model postulated is based upon the following considerations:

-Soliton interactions with life systems have been earlier modelled by Fröhlich and Davydov, and later improved by Pang et al.

-The "pro-life" frequency bands that we identified earlier by an acoustic wave function, show 12 apparent oscillators that we interpret as eigenvalues (also known as scalars and minimum energy values) of coherent like vibrations.

-Numerically similar eigen-frequencies/eigenvalues have earlier been calculated by Ritz for coherent sound patterns of thin membranes.

- We postulate that the calculated discrete EM radiation waves and eigen-frequencies are characteristic for all types of living cells.

-Cellular plasma water exhibits a highly arranged 3-D geometric structure as a liquid crystal, that exhibits surface interactions with macromolecular structures and plays a crucial role in cellular electromagnetism.

- The combination of macromolecules and surrounding water molecules can be conceived as vibrating lattices, sensitive to photon/phonon/soliton interactions.

- Supplying internal and external energy to this matrix induces coherent vibration domains of proteins/DNA/H2O, with a distinct spreading and life-time.

- Such coherent domains can adopt minimal energy levels and can take the form of Bose-Einstein condensates through energy pumping.

- Current quantum biology studies indicate that long range coherency with sufficient life times occur at life temperature ranges.

- The life-sustaining frequencies of our acoustic model can be conceived as solitons having discrete energies, velocity and wave length.

- External solitons can interact with endogenous soliton guided structures in cells through various mechanisms.

- Dark solitons (anti-solitons) can annihilate life-sustaining solitons (by deconstructive interference) and thereby decrease stability of macromolecules.

- Interaction of solitons with BSE condensates and macromolecular lattices can largely prolong their life-times.

- Interaction of solitons/polarons with BSE condensates requires functional trapping and coupling of these energies as modeled by toroidal geometry (see our earlier studies).

- Such toroidal geometry can lead to phase-conjugation, constructive interference or to transmutation of wave modalities that also can lead to regeneration of pro-life solitons by torus scattering.

- Torus mediated coupling of photons, phonons, solitons and BSE condensates is part of a fractal configuration operating at various scales of life organisms.

- It is postulated that solitons, as self-reinforcing solitary waves, show constructive interference and interact with complex biological phenomena such as cellular self-organisation.

## 8. Coherency promoting effects of semiconducting materials

There is an analogy between resonance phenomena in biological cells and resonances in (artificial) semiconductors, while mathematical models about phonon dynamics in semiconducting materials are available (Vasconcellos, 2012, 2013). It has been concluded that the principle of non-equilibrium Fröhlich–Bose–Einstein Condensation like longitudinal-acoustic vibrations (AC phonons) in biological fluids and in typical inorganic semiconductors is quite similar. In case of polar crystalline semiconducting lattices, a macroscopic polarization can be generated by excitations of external electromagnetic waves, which subsequently leads to the emission of coherent THz radiation. For example, the exciton-polaritons, which are bosonic quasiparticles, exist inside microcavities of a semiconductor and consist of a superposition of excitons and cavity photons. Due to their small effective mass Bose–Einstein condensation (BEC) can be realized at temperatures orders higher than that for ultracold atomic gases. The lifetime of polaritons in such semiconductors, that make use of two-dimensional layered materials, are comparable to or even shorter than the thermalization time of 10-100 ps. This property provides them with an inherently non-equilibrium Bose-Einstein nature, which accordingly displays many of the features that would be expected of Bose-Einstein condensates (BECs).

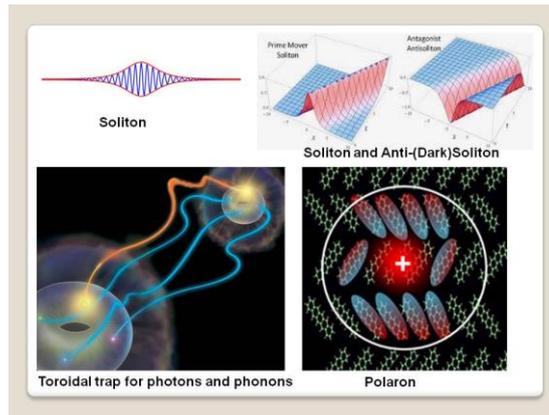

*Figure 6: Solitons (Polarons) and Dark Solitons, formed by toroidal coupling*

Materials used in these types of semiconductors are for example GaAs (Gallium arsenide) and CdTe (Cadmium telluride), which show exciton-polariton condensation at cryogenic temperatures in the vicinity of about 10 K (Byrnes et al., 2014). Typical planar microcavities for exciton-polaritons are applied consisting of several quantum wells sandwiched by distributed Bragg reflectors, see Fig. 7. The quantum wells are thin layers (typically of the order of 10 nm) of a relatively narrow bandgap material (such as GaAs or CdTe) surrounded by a wider bandgap material (doped with Al or Mg respectively).

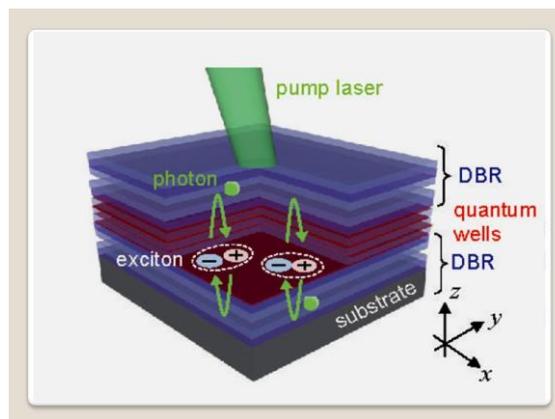

*Figure 7. Exciton-polariton condensation. Typical device structure supporting exciton-polaritons. Excitons, consisting of a bound electron-hole pair, exist within the quantum well layers. These are sandwiched by two distributed Bragg Reflectors (DBRs), made of alternating layers of semiconductors with different refractive indices. The DBRs form a cavity which strongly couples a photon and an exciton to form an exciton-polariton. Polaritons are excited by a pump laser incident from above or by heat (Byrnes et al., 2014)*

The operating temperature of such coherent semiconductors can be further increased from 10 K to room temperature by coupling of the polarons and the polaritons within nano-minerals exhibiting a 3D toroidal-like macroscopic geometry (an equilibrium BEC's will not be possible in a 2D system). Doped silicate minerals are one of the possible building block for these semiconductors (Geesink, 2007). The principle of EM wave transmutation is incorporated in the hypothesis that the quantum state properties of such silicates may have played a role in biological evolution (Coyne, 1985), in the sense that they may have been instrumental in the creation of the first living cells (Ferris, 2005). Hydrous aluminium phyllosilicates are available with variable amounts of alkali-, alkaline-, transition-metals and rare earth elements. The silicates are composed of tetrahedral silicate sheets and octahedral hydroxide sheets and are characterised by two-dimensional sheets of corner sharing $SiO_4$ tetrahedral and/or $AlO_4$ octahedral (see Fig.8). The sheet units have the chemical composition $(Al,Si)_3O_4$. Bonding between the tetrahedral and octahedral sheets requires that the tetrahedral sheet becomes corrugated or twisted, causing a ditrigonal distortion to the hexagonal array, by which the octahedral sheet is flattened, see Fig. 8 (Geesink and Meijer, 2016). This structure enables to increase the coherency of waves in a surrounding of potential detrimental EM wave energy and convert this to more life-sustaining radiation modalities.

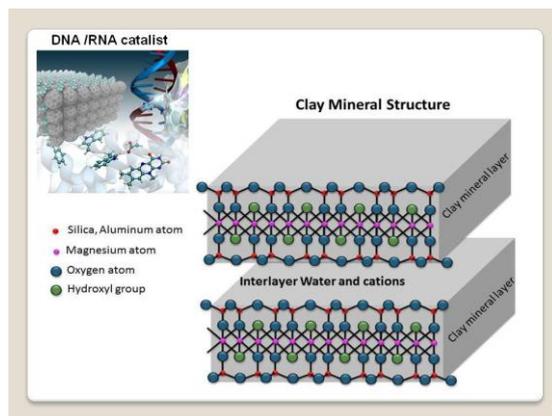

*Figure 8. Example of phyllosilicate-layers, with tetrahedral and octahedral structures*

## 9. Conclusions and final discussion

Solitons, as self-reinforcing solitary waves, are proposed to interact with complex biological phenomena such as cellular self-organisation. The supposed micro-cellular resonances occur at dipolar sites and bond structures of biopolymers, through perturbations of the structure of H-bonding, via quantum states of atoms within pockets of non-localized electrons. The soliton mechanism is seen to be based upon a toroidal mediated electron/phonon coupling of coherent standing waves, on the basis of twelve identified coupled oscillators. The models of Davydov, Fröhlich, Pang, Luzzi, Srobar and Sinkala, discussed in this paper, describe cellular and biomolecular electromagnetism. An addition to these models is proposed, providing a bio-soliton model with a predictive value. The proposed extended model complements the earlier proposed "macroscopic wave function" of the soliton models of Davydov, Fröhlich, and Pang with a coherent frequency scale of an "acoustic wave function", which makes use of twelve coupled oscillators, of which the eigen-frequencies are precisely located in numerical domains in between Hertz and PHz. The complementary aspect of the model has been validated by experimentally observed coherent electromagnetic wave frequencies, as has been practised in 240 published biological studies from 1966 till 2016, in which a spectrum of non-thermal electromagnetic waves were exposed to living cells and intact organisms. The distinct assembly of identified, life supporting EM eigen-frequencies is postulated to represent a coherent electromagnetic field and touches upon the phenomenon of Bose-Einstein condensation at life temperatures. The classical calculation of Ritz concerning acoustic eigen-frequencies, and the experiments of thin vibrating membranes of Chladni are in line with the observed frequency patterns. The calculation of distances of the twelve acoustic tones also complies with a tone scale, of which the position of seven tones of this scale is calculated according a diatonic tone scale as proposed by Pythagoras; the remainder five intermediate tones are calculated using the Pythagorean calculation for flats developed during the Renaissance (Meijer and Geesink, 2016). Measured independent biological radiation data, which show beneficial effects at typical frequencies in the broad Hz, kHz, MHz, GHz, THz and PHz range, are precisely located at eigen-frequencies of the "acoustic wave function" with a mean accuracy of 0.4%. The radiation features showing clear detrimental effects at typical frequencies are located just in between these eigen-frequencies of the "acoustic wave function". It is further considered that the particular "acoustic wave function" may be instrumental in the self-organisation of living cells with their constituting (glyco)-proteins and oligonucleotides (RNA/DNA), which are electromagnetically organized via dedicated quantum states related to the supposed beneficial frequencies. The model and data support the concept of coherent quantized electromagnetic states in living organisms. It is postulated further that detrimental influences of electromagnetic pulses on living cells are initiated by a critical amount of decoherency inducing dark-solitons, exhibiting frequencies, which are not compatible with the pro-life wave characteristics.

Finally, it is envisioned that semiconducting nanomaterials may become available in the near future to produce discrete stabilizing eigen-frequencies characteristic of functional cells, that may prevent possible disturbing influences by de-coherence induction. In potential, even man-made electromagnetic technologies could be further improved in this manner by which anharmonicity is suppressed to obtain a bio-compatible status. Such a technology can be based, for instance, on the principle of toroidal trapping by adding beneficial modes to man-made electromagnetic signals. The latter developments can be envisioned on the basis of the toroidal model as proposed in more detail recently (Meijer and Geesink, 2016).

Zabusky NJ. Interaction of "solitons" in collisionless plasma and the recurrence of initial states, Physical review letters, august, 1965.

Zhadhin M, Barnes F. Frequency and amplitude windows at combined action of DC and low frequency AC magnetic fields on ion thermal motion in a macromolecule: Theoretical analysis. Bioelectromagnetics, 2005; 26(4):323-330.

**Appendix 1: Frequencies reported in biological studies, that *stabilize* cells in vitro or improve in vivo conditions of cells, versus calculated algorithmic frequencies.**

Author, year (w;x;y) >z Hz

Number) Author, year of published biological experiment (applied biological frequency: w; first nearby calculated algorithmic frequency: x, difference in between applied frequency and nearby calculated frequency: y %) > applied frequency normalized to the quantized acoustic scale: z Hz

**Hz-scale**
1) Moore, 1979 (0.3; 0.296; +1.35%) > 307.2 Hz
2) Persinger, 2015 (0.445; 0.444; -0.23%) > 455.7 Hz
3) Persinger, 2015 (0.473; 0.474; -0.21%) > 484.4 Hz
4) Persinger, 2015 (0.482; 0.474; +1.69%) > 493.6 Hz
5) Persinger, 2015 (0.499; 0.500; -0.2%) > 255.5 Hz
6 ) Kole, 2011 (0.6; 0.0595; +0.83%) > 307.2 Hz
7) Fröhlich, F., 2014 (0.750; 0.750; 0.00%) > 384.0 Hz
8) Yu L., 2015 (0.952 Hz; 0.949; +0.29%) > 487.4 Hz
9) Mayrovitz, 2004 (1.000; 1.000; 0.00%) > 256 Hz
10) Hartwich, 2009 (1.000; 1.000; 0.00%) > 256 Hz
11) Sanchez-Vives, 2000 (1.000; 1.000; 0.00%) > 256 Hz
12) De Mattei, 2006 (2.000; 2.000; 0.00%) > 256 Hz
13) Ricci, 2010 (2.000; 2.000; 0.00%) > 256 Hz
14) Hartwich, 2009 (2.000; 2.000; 0.00%) > 256 Hz
15) Hartwich, 2009 (3.200; 3.160; +1.27%) > 409.6 Hz
16) Hartwich, 2009 (4.000; 4.000; 0.00%) > 256 Hz
17) Fadel, 2005 (4.500; 4.500; 0.00%) > 288 Hz
18) Selvam, 2007 (5.000; 5.063; -1.24%) > 320 Hz
19) De Mattei, 2009 (5.000; 5.063; -1.24%) > 320 Hz
20) Sancristóbal, 2014 (6.000; 6.000; 0.00%) > 384 Hz
21) Lisi, 2008 (7.000; 7.100; -1.41%) > 448.0 Hz
22) Ross, 2015 (7.500; 7.590; -1.19%) > 480 Hz
23) Leoci, 2014 (8.000; 8.000; 0.00%) > 256 Hz
24) Belyaev, 1998 (9.000; 9.000; 0.00%) > 288.0 Hz
25) Kole, 2011 (9.4; 9.48; - 0.85%) > 300.8 Hz
26) Golgher, 2007 (10.00; 10.10; -1.00%) > 320 Hz
27) Hood, 1989 (10.00; 10.10; -1.00%) > 320 Hz
28) Fröhlich, F., 2014 (10.00; 10.10; -1.00%) > 320 Hz
29) Kole, 2011 (10.7; 10.66; +0.37%) > 342.4 Hz
30) Golgher, 2007 (13.50; 13.50; 0.00%) > 432 Hz
31) Murray, 1985 (15.00; 15.19; -1.25%) > 480 Hz
32) Lei, 2013 (15.00; 15.19; -1.25%) > 480 Hz
33) Ross, 2015 (15.00; 15.19; -1.25%) > 480 Hz
34) Dutta, 1994 (16.00; 16.00; 0.00%) > 256 Hz
35) Belyaev, 2001 (16.00; 16.00; 0.00%) > 256 Hz
36) Mayrovitz, 2004 (16.00; 16.00; 0.00%) > 256 Hz
37) Hartwich, 2009 (17.10; 16.90; +1.18%) > 273.6 Hz
38) Eusebio, 2012 (20.00; 20.20; -0.99%) > 320 Hz
39) Loschinger, 1999 (20.00; 20.25; - 1.25%) > 320 Hz
40) Chen, 2007 (20.00; 20.25; - 1.25%) > 320 Hz
41) Prato, 2013 (30.0; 30.36; - 1.2%) > 480 Hz
42) Cane, 1993 (37.50; 37.90; -1.06%) > 300 Hz
43) Ceccarelli, 2014 (37.50; 37.90; -1.06%) > 300 Hz
44) Singer, 1999 (40.00; 40.50; -1.24%) > 320 Hz
45) Bastos, 2014 (40.50; 40.50; 0.00%) > 324 Hz
46) Golgher, 2007 (40.50; 40.50; 0.00%) > 324 Hz
47) Reite, 1994 (42.70; 42.70; 0.00%) > 341.6 Hz
48) Blackman, 1985 (45.00; 45.60; -1.32%) > 360 Hz
49) Wei, 2008 (48.00; 48.00; 0.00%) > 384 Hz
50) Cheing, 2014 (50.00; 50.60; 1.19%) > 400 Hz
51) Reale, 2014 (50.00; 50.60; 1.19%) > 400 Hz
52) Segatore, 2014 (50.00; 50.60; 1.19%) > 400 Hz

53) Golgher, 2007 (54.00; 54.00; 0.00%) > 432 Hz
54) Blumenfeld, 2015 (60.00; 60.75; -1.24%) > 480 Hz
55) Braun, 1982 (72.00; 72.00; 0.00%) > 288 Hz
56) Tabrah, 1990 (72.00; 72.00; 0.00%) > 288 Hz
57) Luben RA, 1982 (72; 72; 0%) > 288 Hz
58) Varani, 2002 (75.00; 75.78; -1.01%) > 300 Hz
59) De Mattei, 2006 (75.0; 75.78; -1.01%) > 300 Hz
60) Wang, 2016 (75.0; 75.78; -1.01%) > 300 Hz
61) Veronesi, 2014 (75.0; 75.78; -1.01%) > 300 Hz
62) Reed, 1993 (76.00; 75.90; +0.13%) > 303.6 Hz
63) Singer, 1999 (91.00; 91.18; -0.20%) > 364 Hz
64) Singer, 1999 (100.0; 101.1; -1.09%) > 400 Hz
65) Wen, 2011 (100.0; 101.1; -1.09%) > 400 Hz
66) De Mattei, 2006 (110.0; 108.0; +1.85%) > 440 Hz
67) Douglas, 2001 (120.0; 121.5; -1.24%) > 480 Hz
68) Buhl, 2003 (150.0; 151.6; -1.06%) > 300 Hz
69) Cheron, 2004 (160.0; 162.0; -1.24%) > 320 Hz
70) Chrobak, 2000 (200.0; 202.2; -1.09%) > 400 Hz
71) Schmitz, 2001 (200.0; 202.2; -1.09%) > 400 Hz
72) Kole, 2011 (242.6; 242.9; - 0.12%) > 485.2 Hz
73) Kole, 2011 (274; 269.6; +1.6%) > 274 Hz
74) Mayrovitz, 2004 (300.0; 303.1; -1.02%) > 300 Hz
75) Foffani, 2003 (300.0; 303.1; -1.02%) > 300 Hz

**kHz-scale**
76) Pohl, 1986; (33.00; 32.76; +0.73%) > 257.8 Hz
77) Conner-Kerr, 2015 (35.00; 34.50; +1.45%) > 273.4 Hz
78) Pitt, 2003 (70; 69.0; +1.5%) > 273.4 Hz
79) Hernández-Bule, 2014 (448.0; 442.4; +1.27%) > 437.5 Hz

**MHz-scale**
80) Kyung Shin Kang, 2013 (0.500; 0.498; +0.40%) > 488.3 Hz
81) Kyung Shin Kang, 2013 (1.000; 0.995; +0.50%) > 488.3 Hz
82) Takebe, 2013 (1.000; 0.995; +0.50%) > 488.3 Hz
83) Bandyopadhyay, 2014 (1.000; 0.995; +0.50%) > 488.3 Hz
84) Kyung Shin Kang, 2013 (1.500; 1.490; +0.67%) > 366.2 Hz
85) Takebe, 2013 (3.000; 2.990; +0.33%) > 366.2 Hz
86) Bandyopadhyay, 2014 (3.770; 3.730; +1.07%) > 460.2 Hz
87) Zou, 2007 (5.000; 4.970; 0.60%) >305.2 Hz
88) Pokorny, 2009 (8.000; 7.960; 0.50%) > 488.28 Hz
89 Bandyopadhyay, 2014 (20.00; 19.90; 0.50%) > 305.2 Hz
90) Stolfa, 2007 (21.20; 21.24; 0.19%) > 323.5 Hz

**GHz-scale**
91) Beneduci, 2005 (46.00; 45.79; +0.46%) > 342.7 Hz
92) Fröhlich, ref. G. Schmidl, (46.00; 45.79; +0.46%) > 342.7 Hz
93) Beneduci, 2005 (51.05; 51.54; -0.95%) > 380.4 Hz
94) Fröhlich, ref. G. Schmidl (61.20; 61.09; +0.18%) > 456.0 Hz
95) Radzievsky AA. 2004 (61.22; 61.09; + 0.22%) > 456.1 Hz
96) Kalantaryan, 2010 (64.50; 65.23; -1.12%) > 480.6 Hz
97) Beneduci, 2005 (65.00; 65.23; -0.35%) > 484.3 Hz

**THz-scale**
98) Sukhova, 2007 (0.129; 0.1305; -1.15%) > 480.6 Hz
99) Fedorov, 2011 (2.300; 2.316; -0.69%) > 267.8 Hz
100) Kirichuk, 2013 (3.680; 3.710; -0.81%) > 428.4 Hz

**Optical Nm's scale**
101) Hamblin, 2012 (254; 252.8; +0.48%) > 268.4 Hz
102) Osman and Valadon, 1977 (370; 374; -1.1%) > 368.5 Hz
103) Almeida-Lopes, 2001 (393; 399.5; -1.63%) > 346.9 Hz
104) De Sousa, 2013 (395; 399.5; -1.13%) > 345.1 Hz
105) Gungormus, 2009 (404; 399.5; +1.13%) > 337.5 Hz
106) Rezende, 2007 (415; 420.8; 1.38%) >328.5 Hz
107) Hawkins, 2007 (415; 420.8; -1.38%) >328.5 Hz
108) Bowmaker & Dartnall, 1980 (420.0; 420.8; 0.19%) > 324.6 Hz
109) Hussein, 2011 (445; 449.8; -1.07%) > 306.4 Hz
110) Carotenoid (450; 449.8; +0.1%) >303.0 Hz
111) Reddy, 2003, (452; 449.8; +0.49%) > 301.6 Hz
112) Silveira, 2011 (452; 449.8; +0.49%) > 301.6 Hz
113) Pereira, 2002 (452; 449.8; +0.49%) > 301.6 Hz
114) Chlorophyll b (453; 449.8; +0.7%) > 308.2 Hz
115) Fushimi, 2012 (456; 449.8; + 1.38%) > 299.0 Hz

116) Adamskaya, 2011 (470; 473.4; -0.7%) > 290.1 Hz
117) Cheon, 2013 (470; 473.4; -0.7%) > 290.1 Hz
118) Palacios, 1997 (503.96; 505.6; -0.32%) > 270.5 Hz
119) Palacios, 1997 (505.86; 505.6; +0.05%) > 269.5 Hz
120) Meyer, 2010 (525; 532.5; -1.4%) > 259.7 Hz
121) De Sousa, 2010 (530; 532.5; -0.47%) > 257.3 Hz
122) Weng, 2011 (532; 532.5; -0.09%) > 256.3 Hz
123) Bowmaker & Dartnall, 1980 (534; 532.5; -0.28%) > 255.3 Hz
124) Phycoerythrin (565; 561.0; +0.7%) >482.6 Hz
125) Bowmaker & Dartnall, 1980 (564; 561.0; 0.54%)> 483.4 Hz
126) Weiss, 2005 (590; 599.1; -1.52%) > 462.1 Hz
127) Vojisavljevic, 2007 (596; 599.1; - 0.52%) > 457.6 Hz
128) Komine, 2010 (627; 631.3; -0.68%) > 434.9 Hz
129) Tada, 2009 (629; 631.3; -0.36%) > 433.5 Hz
130) Adamskaya, 2011 (629; 631.3; -0.36%) > 433.5 Hz
131) Karu 2004 (630, 631.3; -0.21%) > 432.8 Hz
132) Huang, 2007 (630; 631.3; -0.21%) > 432.8 Hz
133) 129) Yu, 1997 (630; 631.3; -0.21%) > 432.8 Hz
134) Rabelo, 2006 (632.8; 631.3; +0.24%) >430.9 Hz
135) Carvalho, 2006 (632.8; 631.3; +0.24%) > 430.9 Hz
136) Fahimipour, 2013 (632.8; 631.3; +0.24%) > 430.9 Hz
137) Fushimi, 2012 (638; 631.3: +1.06%) > 427.4 Hz
138) Lacjaková, 2010 (670; 674.0; +0.59%) > 407.0 Hz
139) Lanzafame, 2007 (670; 674.0; +0.59%) > 407.0 Hz
140) Reis, 2008 (670; 674.0; +0.59%) > 407.0 Hz
141) Moore, 2005 (675; 674.0; +0.15%) > 403.9 Hz
142) Chlorophyll a (675; 674.0; +0.14%) > 403.9 Hz
143) Viegas, 2007 (685; 674; 1.6%) > 398.0 Hz
144) Sousa, 2010 (700; 710.1; -1.42%) > 389.5 Hz
145) Choi, 2012 (710; 710.1; -0.01%) > 384.0 Hz
146) Vojisavljevic, 2007 (829; 841.6; - 1.50%) > 328.9 Hz
147) Biscar, 1976 (855.0; 841.6; +1.6%) > 328.9 Hz
148) Hu, 2014 (4300; 4260.0; +0.93%) > 253.6 Hz

## Appendix 2: Frequencies reported in biological studies, that *destabilize* cells in vitro or in vivo, versus calculated algorithmic frequencies

Number) Author, year of published biological experiment (applied biological frequency: w; first nearby calculated algorithmic frequency: x, difference in between applied frequency and nearby calculated frequency: y %) > applied frequency normalized to the quantized acoustic scale: z Hz

**Hz-scale**
1) Puharich (6.600; 6.540; +0.92%) > 422.4 Hz
2) Pfluger, 1996 (16.7, 16.43; +1.64%) > 267.2 Hz
3) Ghione, 1996 (37; 37.0; -0.1%) > 296.0 Hz
4) Ahmed, 2013 (200; 197.2; +1.4%) > 400 Hz
5) Ahmed, 2013 (250; 249.5; +0.2%) > 250 Hz
7) Ahmed, 2013 (350; 353; -0.86%) > 350 Hz
8) Ahmed, 2013 (400; 394.3; +1.4%) > 400 Hz
9) Ahmed, 2013 (500; 499; +0.2%) > 250 Hz

**KHz-scale**
10) Sausbier, 2004 (5, 5.018; -0.36%) > 312.5 Hz

**MHz-scale**
11) Brown-Woodman, 1988 (27.12; 27.41; - 1.1%) > 413.82 Hz
12) Tofani, 1986 (27.12; 27.41; - 1.1%) > 413.82 Hz
13) Bawin, 1973 (147; 146.2; +0.55%) > 280.38 Hz
14) Jauchem and Frei 1997 (350; 348.8; -0.3%) > 333.78 Hz
15) Sanders, 1980, (591; 584.9; +1.1%) > 281.8 Hz
16) De Pomerai, 2000, Daniells, 1998 (750; 740.3; +1.3%) > 357.6 Hz
17) Maskey, 2014, (835; 826.9; + 0.98%) > 398.21 Hz
18) Donnellan, 1997 (835; 826.9; +0.98%) > 398.2 Hz
19) Adey, 2006 (836; 826.9; + 1.1%) > 398.6 Hz
20) Mashevich, 2003 (830; 826.9; +0.38%); 395.8 Hz
21) Luukkonen, 2009 (872; 877.12; -0.58%) > 415.8 Hz
22) Hao, 2012 (916; 930.28; -1.5%) > 436.8 Hz
23) Johnson and Guy, 1972 (918; 930.3; -1.3%) > 437.7 Hz
24) Zmyslony, 2004 (930; 930.3; -0.03%) > 443.5 Hz
25) Maes, 1997 (935.2; 930.3; +0.5%) > 445.9 Hz

26) Jauchem and Frei 2000 (1000; 986.9; +1.3%) > 476.8 Hz
27) De Pomerai, 2003 (1000; 986.9; +1.3%) > 476.81 Hz
28) Lu, 1992 (1250; 1239.8; +0.82%) > 298.0 Hz
29) Oscar and Hawkins, 1977 (1300; 1315.4; -1.1%) > 310.0 Hz
30) Wake, 2007 (1500; 1480.6; +1.3%) > 357.6 Hz
31) Schirmacher, 1999 (1750; 1754.2; - 0.24%) > 417.2 Hz
32) Iyama, 2004 (2000; 1973.8; +1.3%) > 476.8 Hz
33) Senavirathna, 2014 (2000; 1973.8; +1.3%) > 476.8 Hz
34) Aydogan, 2015 (2100; 2093.0; +0.33%) > 250.3 Hz
35) Grin AN, 1974, 2375, 2339.5; +1.5%) > 283.1 Hz
36) Shandalal, 1979 (2375; 2339.5; +1.5%) > 283.1 Hz
37) Lai et al, 1987, 1988 (2450; 2479.7; -1.2%) > 292.1 Hz
38) Deshmukh, 2015 (2450; 2479.7; -1.2%) > 292.1 Hz
39) Switzer and Mitchell, 1977 (2450; 2479.7; -1.2%) > 292.1 Hz
40) Kesari KK, 2010 (2450; 2479.7; -1.2%) > 292.1 Hz
41) Shokri S. 2015 (2450; 2479.7; -1.2%) > 292.1 Hz
42) FigueiredoI, 2004 (2500; 2479.7; +0.8%) > 298.0 Hz
43) Bin, 2014 (2576; 2630.7; - 2.1%) > 307.1 Hz
44) Thomas et al. 1982 (2800; 2790.1; +0.35%) > 333.8 Hz
45) Albert EN, 1977 (2800; 2790.1; +0.35%) > 333.8 Hz
46) Frei and Jauchem, 1989 (2800; 2790.1; +0.35%) > 333.78 Hz
47) Gandhi, CR., 1989 (2800; 2790.1; +0.35%) > 333.8 Hz
48) Siekierzynski, 1972 (2950; 2961.1; - 0.38%) > 351.7 Hz
49) Pu, 1997 (3000; 2961.1; +1.3%) > 357.61 Hz
50) Grodon, 1970 (3000; 2961.1; +1.3%) > 357.61 Hz
51) Tolgskaya, 1973 (3000; 2961.1; +1.3%) > 357.6 Hz
52) DÁndrea et al, 1994 (5600; 5580.2; +0.35%) > 333.8 Hz
53) Jensh, 1984 (6000; 5922.2; +1.3%) > 357.6 Hz
54) Copty. 2006 (8500; 8371.8; +1.5%) > 253.31 Hz
55) Goldstein and Sisko, 1974 (9300; 9358; -0.6%) > 277.2
56) Zhang Y, 2014, (9417; 9358.1; +0.63%) > 280.7 Hz
57) Zhang, 2014 (9410; 9358.1; +0.56%); 280.4 Hz

**GHz-scale**
58) Sharma, 2014 (10.0; 9.92; +0.8%); 298.0 Hz
59) Jauchem and Frei 2000 (10.0; 9.92; +0.8%) > 298.0 Hz
60) FigueiredoI, (2004, 10.5; 10.53; -0.2%) > 312.9 Hz
61) Porcelli, (10.40; 10.52; -1.16%) > 309.9 Hz
62) Paulraj, 2012 (16.50; 16.74; -1.46%); 245.9 Hz
63) Millenbaugh NJ, 2008 (35; 35.29; - 0.82%) > 260.8 Hz
64) Roza K. 2010 (35; 35.29; - 0.82%) > 260.8 Hz
65) Shock, 1995 (35; 35.29; - 0.82%) > 260.8 Hz
66) Kesari KK, 2009 (50; 50.25, -0.50%) > 372.5 Hz
67) Tafforeau M, 2004 (105; 105.85; - 0.80%) > 391.2 Hz

**THz-scale**
68) Kirchuk, 2008 (0.24; 0.238; -0.78%) > 447.0 Hz
69) Webb and Dodds, 1968 (0.136, 0.134; + 1.54%) > 253.3 Hz
70) Wilmink, 2010 (2.52; 2.54; - 0.78%) > 293.4 Hz
71) Homenko, 2009 (0.1; 0.1; -0.5%) > 372.5 Hz

**Optic nm's scale**
72) Kitchel E., 2000 (435; 434.7; -0.1%) > 313.4 Hz
73) Glazer-Hockstein, 2006, Algvere PV, 2006, Marshall J, 2006, Tomany S.C, 2004, 2008,
Smick K, 2013 (435.0; 435.1; -0.02%) > 313.4 Hz
74) Ham, 1980 (441.0; 435.1; -1.4%) > 309.1 Hz
75) Hu, J., 2014 (466.0; 461.4; +1.0%) > 292.5 Hz
76) Ueda, T., 2009 (465.0; 461.4; + 0.8%) > 293.2 Hz
77) Moore, P, 2005 (810; 820.2; -1.2%) > 336.6 Hz

**Appendix 3: The different beneficial and detrimental health effects reported in biological studies caused by electromagnetic fields**

114 different frequencies of electromagnetic waves in 145 independent published biological studies could be found, which show beneficial effects on cells in vivo and in vitro. The experiments are in the areas of: neuro-stimulation, brain stimulation, spinal cord stimulation, self-assembly via tunnelling current of microtubulins, transcranial magnetic stimulation, reduction of Parkinson, anti-proliferative effects on tumour cells, inhibition of tumour growth, reduced and repression of tumour growth, rhythmic neuronal synchronization, improvement of memory, improvement of attention, wound healing, decrease of inflammatory cells, increase of bone growth, reduction of diabetic peripheral neuropathy, increase of fibroblast proliferation, stimulation of angiogenesis, granulation of tissue formation and synthesis of collagen, promotion of proliferation of human mesenchymal stem cells, entorhinal-hippocampal interactions, and others (references of the biological studies can be obtained upon request, authors and publication year have been mentioned in Appendix 1).

60 different measured frequencies of electromagnetic waves could be found in literature, that destabilize living cells in 74 independent biological studies reporting detrimental effects of EM radiation on cells in vivo and in vitro. The experiments in these studies are in the areas of: tumour growth, influence on teratogenic potential, DNA single-strand breaks, gene expression, chromosomal instability, inhibition of cell growth, influences on sperm parameters, influence on sleeping, influence on the permeability of the blood-brain barrier, influence on behaviour, cognitive impairment, learning and memory alterations, maculopathy, influence on specific brain rhythms, alter of protein conformation, effects on blood pressure, cardiovascular responses, phototoxic effects on human eye health, and on the retina, influence on alkaline phosphatase activity and antigen-antibody interaction, cardiovascular effects and others (references can be obtained upon request, authors and publication year have been mentioned in Appendix 2).